\documentclass[a4paper]{article}

\usepackage[english]{babel}
\usepackage[utf8x]{inputenc}
\usepackage[T1]{fontenc}
\usepackage{authblk}
\usepackage{amsmath}
\usepackage{amsfonts}
\usepackage{subfigure}
\usepackage{cite}
\usepackage{atbegshi}
\AtBeginDocument{\AtBeginShipoutNext{\AtBeginShipoutDiscard}}
\usepackage[a4paper,top=3cm,bottom=2cm,left=3cm,right=3cm,marginparwidth=1.75cm]{geometry}

\usepackage{amsmath}
\usepackage{graphicx}
\usepackage[colorinlistoftodos]{todonotes}
\usepackage[colorlinks=true, allcolors=blue]{hyperref}

\title{L\' evy flight movements prevent extinctions and 
maximize population abundances in fragile Lotka-Volterra systems}

\begin{document}

\date{}

\author[a,b,c,d]{Teodoro Dannemann}
\author[d,e,1]{Denis Boyer} 
\author[d,e,f]{Octavio Miramontes}

\begin{flushleft}
\footnotesize 
\affil[a]{ \footnotesize Instituto de Conservaci\'on, Biodiversidad y Territorio, Facultad de Ciencias Forestales y Recursos Naturales, Universidad Austral de Chile} \quad
\affil[b]{Departamento de Ecolog\'ia, Facultad de Ciencias Biol\'ogicas, Pontificia Universidad Cat\'olica de Chile} \quad
\affil[c]{Instituto de Ecolog\'ia y Biodiversidad (IEB), Las Palmeras 1345, Macul, Santiago, Chile.} \quad
\affil[d]{Instituto de F\'\i sica, Universidad Nacional Aut\'onoma de M\'exico, D.F. 04510, M\'exico} \quad
\affil[e]{Centro de Ciencias de la Complejidad, Universidad Nacional 
Aut\'onoma de M\'exico, D.F. 04510, M\'exico}
\affil[f]{Departamento de Matem\'{a}ticas Aplicadas, 
ETSI Aeron\'{a}uticos, Universidad Polit\'{e}cnica de Madrid, Madrid,
Spain.}
\end{flushleft}





\maketitle

\begin{abstract}
 Multiple-scale mobility is ubiquitous in nature and has become instrumental for understanding and modeling animal foraging behavior. However, the impact of individual movements on the long term stability of populations remains largely unexplored. We analyze deterministic and stochastic Lotka-Volterra systems where mobile predators consume scarce resources (prey) confined into patches. In fragile systems, that is, those unfavorable to species coexistence, the predator species has a maximized abundance and is resilient to degraded prey conditions when individual mobility is multiple-scaled. Within the L\'evy flight model, highly superdiffusive foragers rarely encounter prey patches and go extinct, whereas normally diffusing foragers tend to proliferate within patches, causing extinctions by overexploitation. L\'evy flights of intermediate index allow a sustainable balance between patch exploitation and regeneration, over wide ranges of demographic rates. Our analytical and simulated results
can explain field observations and
suggest that scale-free random movements are an important mechanism by which entire populations adapt
to scarcity in fragmented ecosystems.
\end{abstract}

\vspace{1.5cm}


Species extinction, population loss and biodiversity decline represent real dangers for the continuity of life on earth \cite{valiente}. Current extinction rates are several orders of magnitude above normal background rates \cite{pimm2006human, Barnosky, DeVos}. Halting and reversing this trend is a formidable challenge that requires a better understanding of how ecosystems operate, and where interdisciplinary approaches can play an essential role. Over the years, physical and mathematical concepts have provided valuable tools for studying a range of ecological phenomena such as nonlinear and chaotic dynamics in population biology \cite{may}, non-equilibrium phase transitions \cite{solebascompte} or the structure and resilience of ecological networks \cite{pascual}.

Fragile ecosystems are often fragmented, namely, composed of populations separated in space, either because of a natural tendency of individuals to aggregate in patches, or because of human perturbations \cite{levins,Taylor,Ritchie}. Within small areas, populations are more exposed to local extinctions due to demographic stochasticity, or when the growth of an invasive species leads to the overexploitation of slowly recovering resources \cite{Hamilton,Hamilton2,Okubo}. In systems of fragments (metapopulations), the ability of the organisms to move from one fragment to another has been identified as a crucial stabilizing factor that can prevent irreversible decline \cite{levins,hanski,niebuhr2015survival}. 

Interacting species in uniform \cite{may-miramontes,miramontes-rohani,Tainaka,Matsuda,Mobilia} or fragmented \cite{solebascompte,Okubo,maritan} landscapes have been extensively explored with Lotka-Volterra models, a paradigmatic framework in population dynamics \cite{Lotka, Volterra, Bashan, Perhar}. Individual mobility is a key aspect in this approach, and it is usually modeled by standard random walks without long-range displacements (but see \cite{hanert}). In recent years, thanks to the improvement of tracking devices, data analysis have yet revealed that single animal trajectories often contain multiple characteristic scales, calling for new theories and models of mobility beyond simple diffusion \cite{morales,nathan,revilla2008individual,intermsearch,benhamou}. 

A body of recent observations in a variety of context and animal taxa have reported evidence of 
multiple-scale mobility patterns well described by L\'evy flights or L\'evy walks
\cite{gandhibook, bartu2003,RF,boyermonos,Atk,reynolds,Brown,Sims,deJager,mirplosone,humphries}. A widely discussed interpretation, the L\'evy foraging hypothesis, is based on the fact that L\'evy movements represent an efficient random search strategy in unpredictable environments where prey are scarce and distributed in patches\cite{gandhibook,Bar3,bartulevin}. Given a predator having no information on the location of prey, its efficiency (rate of prey capture) is maximized if the forager performs a L\'evy walk with exponent $\beta\approx 2$, a value observed in the field \cite{bartu2003,reynolds,RF,Sims}. Despite the fact that other interpretations exist \cite{boyermonos,Atk}, few studies have discussed the consequences of L\'evy mobility on collective aspects in systems of interacting individuals \cite{prl2013}. We are interested in addressing this question, in particular how entire populations respond in front of drastic changes in resource availability. Remarkably, L\'evy walks have been shown to be evolutionary stable in mussels colonies, by achieving a compromise between reducing the risk of predation and minimizing intra-specific competition for food \cite{deJager}. But in many cases, a seemingly optimal individual foraging strategy may lead to severe resource depletion due to feedback effects \cite{Boyer2,bhat2017does}. The movement strategies considered as efficient for a single individual immersed in a sea of static prey (a common theoretical setup) need to be re-examined for large populations and longer time-scales.

Here, we show by means of previously unexplored analytical arguments and computer simulations that multi-scaled random walks have a significant impact on the stability of metapopulations close to extinction thresholds. We consider both deterministic and stochastic lattice Lotka-Volterra (LV) models \cite{Tainaka,Matsuda,Mobilia}, where the resources are fragmented into areas distant from each other and predators can perform L\'evy flights instead of nearest-neighbor (NN) random walks.

\section*{Analytical population model in patchy landscapes}

We start with a solvable rate equation model defined on a two-dimensional ($2$D) space, where prey are restricted to occupy patches and predators diffuse according to a power-law mobility kernel. Space is made of a regular lattice of $N$ square cells, each of length  
$R_0$. Some cells can contain prey and thus represent \lq\lq patches\rq\rq\  of area $R_0^2$. These patches form a periodic square array for simplicity, with separation distance between neighboring patches given by $l_0R_0$, where $l_0>1$ is an integer. No prey can be present outside of the patches. The predator and prey densities in cell ${\mathbf n}$ at time $t$, where ${\mathbf n}\in \mathbb{Z}^2$, are denoted as $a_{\mathbf n}(t)$ and $b_{\mathbf n}(t)$, respectively. Outside of the prey patches, $b_{\mathbf n}(t)=0$ but $a_{\mathbf n}(t)$ can be $\ne 0$. Assuming that occupied cells contain many individuals and fluctuations are negligible, we write the Lotka-Volterra equations \cite{Mobilia}:
\begin{eqnarray}
\frac{da_{\mathbf n}}{dt}&=&
-\lambda_0 a_{\mathbf n}+\lambda_0\sum_{{\boldsymbol \ell}}
P({\boldsymbol \ell})a_{{\mathbf n}-{\boldsymbol \ell}}+
\lambda a_{{\mathbf n}}b_{{\mathbf n}} -\mu a_{\mathbf n} \label{ath}\\
\frac{d b_{\mathbf n}}{d t}&=&\sigma b_{\mathbf 
n}\left(1-\frac{b_{\mathbf n}+a_{\mathbf n}}{K}\right)-\lambda'a_{\mathbf n} b_{\mathbf n}, \label{bth}
\end{eqnarray} 
where $\lambda_0$, $\lambda$, $\mu$ and $\lambda'$ are the predator movement, reproduction, mortality and predation rates, respectively. $K$ is the patch carrying capacity and $\sigma$ the prey reproduction rate. The cell-to-cell predator jump distribution $P({\boldsymbol \ell})=P(\ell_x,\ell_y)$ is given for simplicity by the product of two one-dimensional scale-free distributions with integer argument and exponent $\beta>1$:
\begin{equation}
P({\boldsymbol \ell})=p(\ell_x)p(\ell_y)\quad {\rm with}\quad p(\ell)=p_0\delta_{\ell,0}+(1-p_0) f(\ell), 
\end{equation}
where  $f(0)=0$, $f(\ell)=|\ell|^{-\beta}/[2\zeta(\beta)]$ for $\ell=\pm1,\pm2,...$, $\zeta(\beta)=\sum_{n=1}^{\infty}n^{-\beta}$ is the normalization constant and $\delta_{\ell,0}=1$ or $0$ the Kronecker symbol. We use the product of two power-laws because of the lattice symmetry, but other choices lead to similar results (see next Section). The foragers are normally diffusive (Brownian) for $\beta>3$ and super-diffusive (L\'evy) for $1<\beta<3$. In the case $\beta\rightarrow 1$, extremely long steps are taken, which is equivalent in practice to random relocations in space.

The quantity $p_0^2$ represents the probability that a predator remains in the same cell after a movement step, when the latter is too small to bring the predator outside of its current cell. Approximate arguments allow to relate $p_0$ to the patch size: one assumes that predators actually perform continuous steps, inside or across patches, of minimal length $x_0$ which is set to unity in the following. For patches with $R_0>1$, one obtains $p_0=1-1/R_0+(1-R_0^{2-\beta})/[(2-\beta)R_0]$, see SI text.

In the absence of movement ($\lambda_0=0$, or $p_0=1$), the prey and predator abundances are zero at large times everywhere except in prey cells, where Eqs. (\ref{ath}) and (\ref{bth}) reduce to two ordinary differential equations (ODE) for a single patch. They admit two simple stationary fixed points, $(a_0^{(o)},b_0^{(o)})=(0,0)$ and $(a_0^{(u)},b_0^{(u)})=(0,K)$, corresponding to total extinction (by over-exploitation) and predator extinction (by under-exploitation), respectively. A third, globally stable, coexistence fixed point exists for $\mu/\lambda<K$
\cite{Mobilia}: 
\begin{equation}\label{anomove}
a^{(no\ move)}_0=(K-\mu/\lambda)/(1+\lambda'K/\sigma)
\end{equation}
and $b^{(no\ move)}_0=\mu/\lambda$. 
If $K<K_c=\mu/\lambda$, predators go extinct and $b^{(no\ move)}_0=K$. Oscillatory solutions do not exist \cite{Mobilia}.

\begin{figure}
\centering
\includegraphics[scale=0.5]{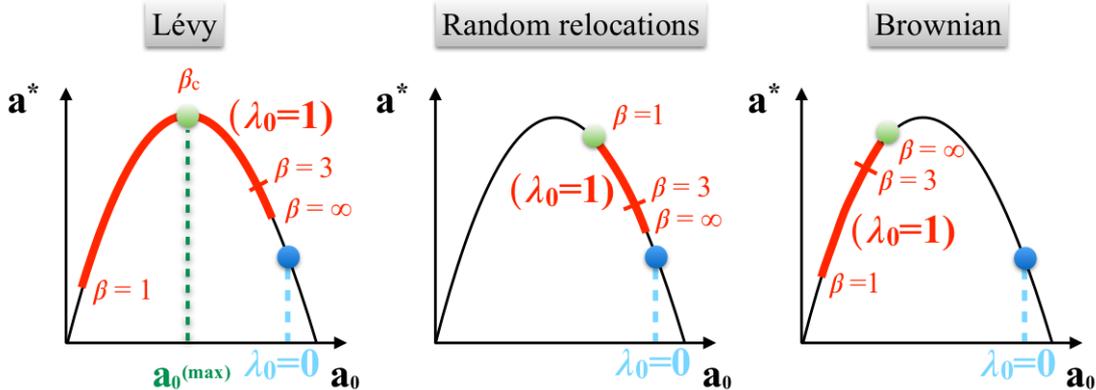}
\caption{When immobile predators ($\lambda_0=0$) over-exploit prey patches ($a_0^{(no\ move)}>a_0^{(max)}$), incorporating mobility ($\lambda_0>0$) usually increases the total predator abundance $a^*$. The strategy maximizing $a^*$ (green disk) can be L\'evy, random or Brownian. The L\'evy strategy is an advantageous response in the most fragile systems, since, there, $a^*$ may otherwise reach low values from two sides. For $\beta>3$ the foragers practically perform NN random walks ($\beta=\infty$ limit) and $a^*$ vary little.}
\label{regimes}
\end{figure}

With mobile individuals ($\lambda_0>0$), the cells are no longer isolated. A quantity of particular interest in this case is the spatially averaged number of predators per unit area,  $a^*\equiv\sum_{\mathbf n} a_{\mathbf n}/N$. We look for non-zero stationary solutions of Eqs. (\ref{ath})-(\ref{bth}). The steady state $a^*$ takes the form (see SI text):
\begin{equation}\label{Np}
a^*=\frac{\lambda}{\mu l_0^2}a_0[K-a_0(1+K\lambda'/\sigma)] 
\end{equation}
with $a_0$ the predator density in the prey patches:
\begin{equation}\label{a0}
a_0=\frac{1}{1+\frac{K\lambda'}{\sigma}}
\left[K-\left(\frac{\lambda}{(2\pi)^2}\sum_{\mathbf n} \int_{\cal B}d{\mathbf k}
\frac{\cos(l_0{\mathbf k}\cdot{\mathbf n})}
{\lambda_0[1-\hat{P}(\mathbf k)]+\mu}\right)^{-1} \right],
\end{equation}
where $\hat{P}(\mathbf k)\equiv\sum_{\boldsymbol \ell} P({\boldsymbol \ell})
e^{-i{\mathbf k}\cdot{\boldsymbol \ell}}$ is the Fourier transform of $P$ and ${\cal B}$ the first Brillouin zone, defined by $-\pi<k_x,k_y<\pi$.

Notably, in Eq. (\ref{Np}) the mean predator density $a^*$ for the whole system obeys a logistic relation with respect to $a_0$, the predator density in {\it one} prey patch (Figure \ref{regimes}). Thus $a^*$ is maximal when $a_0=a^{(max)}_0\equiv K/[2(1+K\lambda'/\sigma)]$, and vanishes at $a_0=0$ and $a_0=2a^{(max)}_0$. (At $2a_0^{(max)}$ and above, the only acceptable stationary solution is $a_0=a_{\mathbf n}=0$.) In the low density regime, $0<a_0<a^{(max)}_0$, predators under-exploit prey: any increase in $a_0$ produces an increase of $a^*$. Whereas at high density, $a^{(max)}_0<a_0<2a^{(max)}_0$, foragers over-exploit the patches: any increase in $a_0$ decreases the total abundance. 
The demographic parameter being fixed, the largest $a_0$ is always obtained in the absence of movement ($\lambda_0=0$). Therefore, some amount of movement will be beneficial (increase $a^*$) if $a^{(no\ move)}_0$ is located in the over-exploitation regime, $a^{(no\ move)}_0>a^{(max)}_0$, implying $\mu<\frac{K\lambda}{2}$. We set this condition in the following, as it is relevant to fragile systems.

We define the optimal movement strategy as the one maximizing the predator abundance $a^*$. Keeping all the parameters fixed except $\beta$, the density $a_0$ given by Eq. (\ref{a0}) can be varied, giving rise to three possibilities. {\it (i)} $a_0=a^{(max)}_0$ for an exponent $\beta_c$ such that $1<\beta_c<3$, see  Figure \ref{regimes}-Left (where $\lambda_0=1$ without loss of generality). The value $\beta_c$ satisfies:
\begin{equation}\label{opt}
\frac{1}{(2\pi)^2}\sum_{\mathbf n}\int_{\cal B}d{\mathbf k}\frac{\cos(l_0{\mathbf k}\cdot{\mathbf n})}
{1-\hat{P}({\mathbf k})+\mu^*}=\frac{2}{K\lambda^*},
\end{equation}
with $\mu^*=\frac{\mu}{\lambda_0}$ and $\lambda^*=\frac{\lambda}{\lambda_0}$.
Recall that the dependence in $\beta$ is contained in the term $\hat{P}({\mathbf k})$. {\it (ii)} If Eq. (\ref{opt}) does not admit any solutions in the interval $(1,3)$, $a^*$ may still reach a maximum for the lowest possible value $\beta=1$, the movement mode that least over-exploits resources (Fig. \ref{regimes}-Middle). {\it (iii)} In the third case, Figure \ref{regimes}-Right, Brownian movement ($\beta\ge3$) provides the optimal strategy, namely, the best way of exploiting in conditions of under-exploitation.

\begin{figure}
\centering
\includegraphics[scale=0.7]{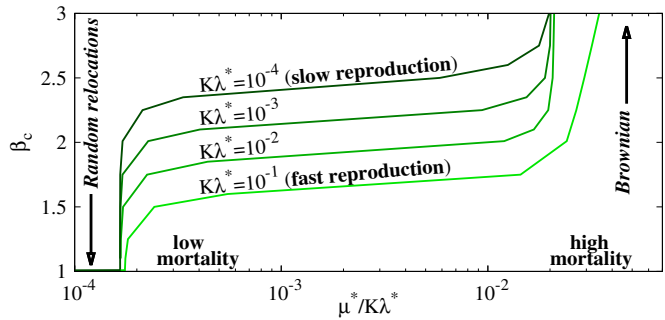}
\caption{L\'evy exponent maximizing predator abundance as a 
function of the reduced mortality rate, for various
reproduction rates $\lambda$, as given by Eq. (\ref{opt}). $R_0=30$ and the patch volume fraction is $4\ 10^{-2}$ ($l_0=5$). At very large (low) $\mu$ the optimal strategy is Brownian (with random relocations or $\beta=1$, respectively). Fast (slow) predator reproduction favors more (less) superdiffusive strategies.}
\label{ranges}
\end{figure}

We explore a realistic ecological situation, where predators are mobile, slowly reproducing and long-lived, {\it i.e.}, $1\gg K\lambda^*\gg \mu^*$ (note: all the demographic rates may be scaled by the movement scale $\lambda_0$). For an environment of low patch density, Figure \ref{ranges} shows that the optimal $\beta_c$ obtained from solving Eq. (\ref{opt}) can be in the L\'evy range $1<\beta_c<3$ and depends little on $\mu^*$ and $K\lambda^*$ over wide intervals. For a fixed predator reproduction rate $K\lambda^*$, the strategy leading to the largest $a^*$ rapidly switch to Brownian or to random relocations at very high and low $\mu^*$, respectively: High predator mortality rates reduce prey over-exploitation and promote Brownian strategies (predators stay close to the prey patch where they were born). Random relocations, in contrast, allow patch regeneration if predators are long-lived. Similarly, low values of the predator reproduction rate $K\lambda^*$ reduce the predation pressure and slowly move $\beta_c$ upward, toward Brownian motion. In the following, we drop the superscripts $^*$ and set the movement rate to $\lambda_0=1$.

\section*{Stochastic lattice Lotka-Volterra model (SLLVM)}

The foregoing analytical results show the importance of L\'evy movements at the population level. What's more, Figure \ref{regimes}-Left allows us to clarify the notion of \lq\lq fragility\rq\rq\ from a movement ecology point of view: a system is most fragile when markedly different ranging modes (here, $\beta\ge3$ and $\beta\rightarrow 1$) bring the system close to distinct zero-abundance fixed points ($a_0^{(o)}$ and $a_0^{(u)}$ above). We focus below on this generic situation and proceed to verify our predictions with simulations in a few representative numerical examples. We also incorporate the effects of fluctuations in the description by building a stochastic model inspired in ref. \cite{Mobilia}.

\subsection*{Rules}

Space is a two-dimensional lattice of $L\times L$ sites of unit area, with periodic boundary conditions. Each site can be empty ($\emptyset$), with a predator ($A$), with a predator reproducing ($AA$) or with a prey ($B$). Double occupation of a site is forbidden (except for the $AA$ reproductive state). The prey is confined to limited areas: $n$ circular patches of radius $R$ are randomly distributed, inside which the sites are initially set to state $B$. (We choose $\pi R^2=R_0^2$ so that patches have the same area than in the analytical model.) Prey cannot occupy sites which are outside of the patches. Monte Carlo simulations are performed over many landscapes with $L^2/5$ initial predators (other numbers do not affect the results).

At each elementary step, an occupied site is chosen randomly and updated as follows:
\begin{itemize}
\item {\it Predator death:} If a predator is selected, it dies with probability $\mu$.
\item {\it Predator movement and reproduction:} A selected surviving predator randomly chooses a site at a distance $\ell$ where $\ell>1$ is drawn from a power-law distribution $P(\ell)=c\ell^{-\beta}$, with $\beta$ an exponent and $c$ the normalization constant. If another predator is present at the new position, the selected predator does not move, otherwise it occupies the new site (only one predator moves at a time). If a prey is present there, the predator eats it and reproduces.
\item {\it Prey reproduction:} If a prey is selected, one of its NN sites (within the patch) is chosen randomly. If that site is empty, a prey offspring is produced there with probability $\sigma$. In other cases nothing happens. 
\end{itemize} 

In these rules, $\lambda'=\lambda=1$ and $K=1$. In the SLLVM of ref. \cite{Mobilia}, all agents were mobile with NN hopping and the carrying capacity was uniform.  Here, prey are static and $K=1$ for the sites belonging to the patches ($K=0$ elsewhere). The fraction of area covered by the patches is ${\cal A}\simeq n\pi R^2/L^2$ (${\cal A}=1/l_0^2$ in the analytic model). A mean-field (MF) solution of our SLLVM can be obtained when the predators are well mixed in the system, {\it i.e.}, in the random relocations regime or $\beta$ close to 1. Neglecting spatio-temporal fluctuations, we show in the SI text that the predator abundance $a^{(MF)}$ is given by Eq. (\ref{anomove}) but with $\lambda$ substituted by ${\cal A}\lambda$.  

\begin{figure}
\centering
\subfigure[$\ \beta\approx 1$]{
\includegraphics[scale=0.39]{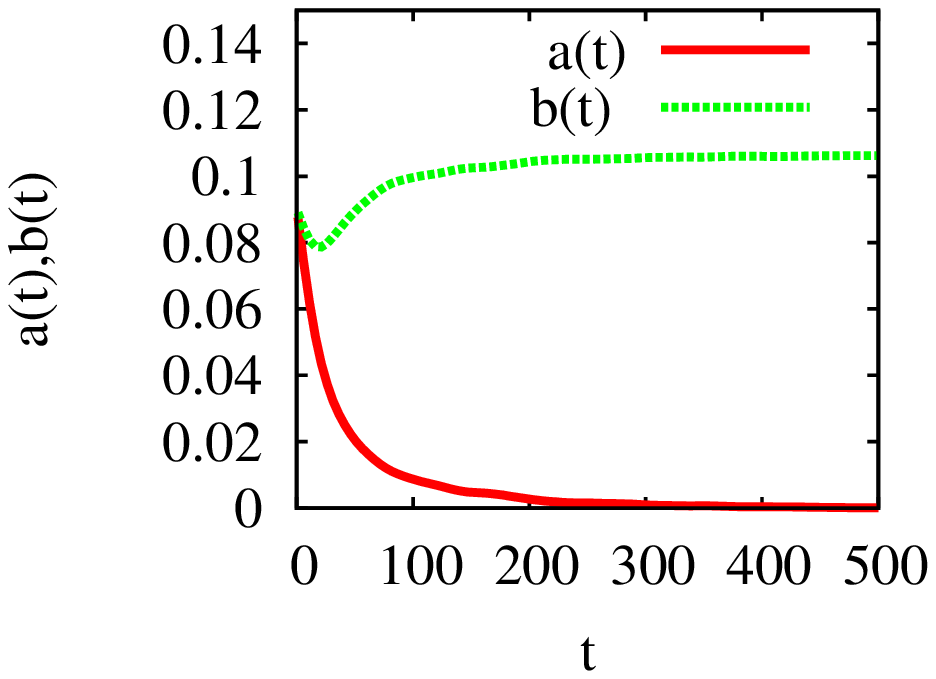}
}
\subfigure[$\ \beta= 3$]{
\includegraphics[scale=0.39]{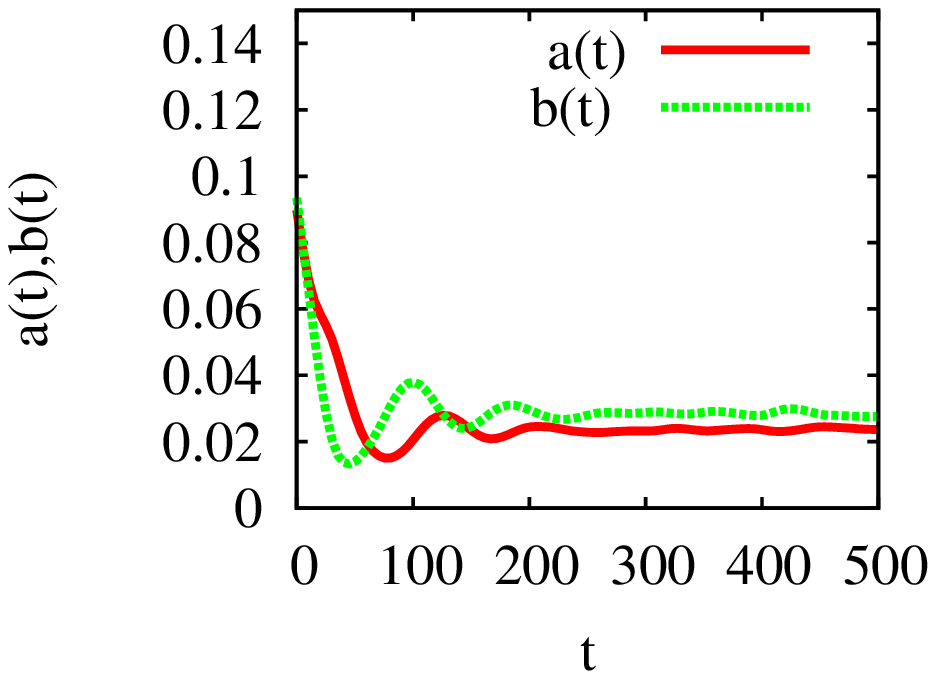}
}
\caption{Evolution of $a(t)$ and $b(t)$ toward quasi-stationary states for a single run, in the case of highly-superdiffusive ($\beta\simeq 1$, left) and Brownian (right) predators, the other parameters are $\mu=0.22$, $\sigma=0.5$, $L=500$, $n=20$. A Monte Carlo time step corresponds to selecting all individuals once on average.}
\label{brownbalcm}
\end{figure}

%
We consider two scenarios. In the first one, prey are scarce and ${\cal A}\ll 1$, such that predators go extinct in the above MF approximation, {\it i.e.}, ${\cal A}\lambda<\mu$. Given a predator mortality rate $\mu$, we choose ${\cal A}=\mu/2$, which is achieved by setting the patch radius to
\begin{equation}\label{R}
R=\sqrt{\mu/(2\pi n)}L,
\end{equation}
where $n$ is fixed. In the second scenario, the predator mortality rate $\mu$ is held fixed and prey abundance varied through the parameters $n$ and $R$.

\begin{figure}
\centering
\subfigure[ ]{
\includegraphics[scale=0.3]{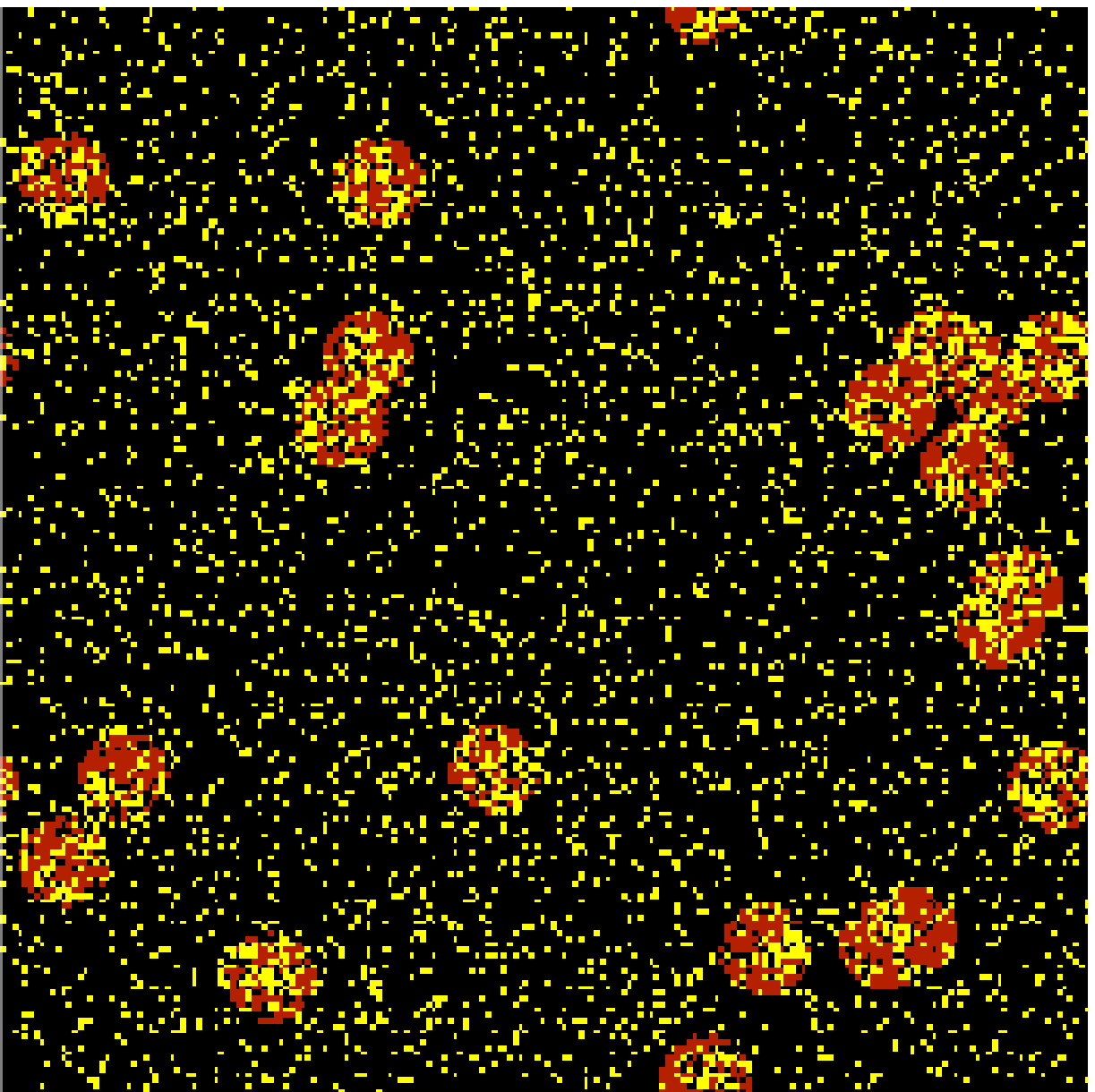} \includegraphics[scale=0.3]{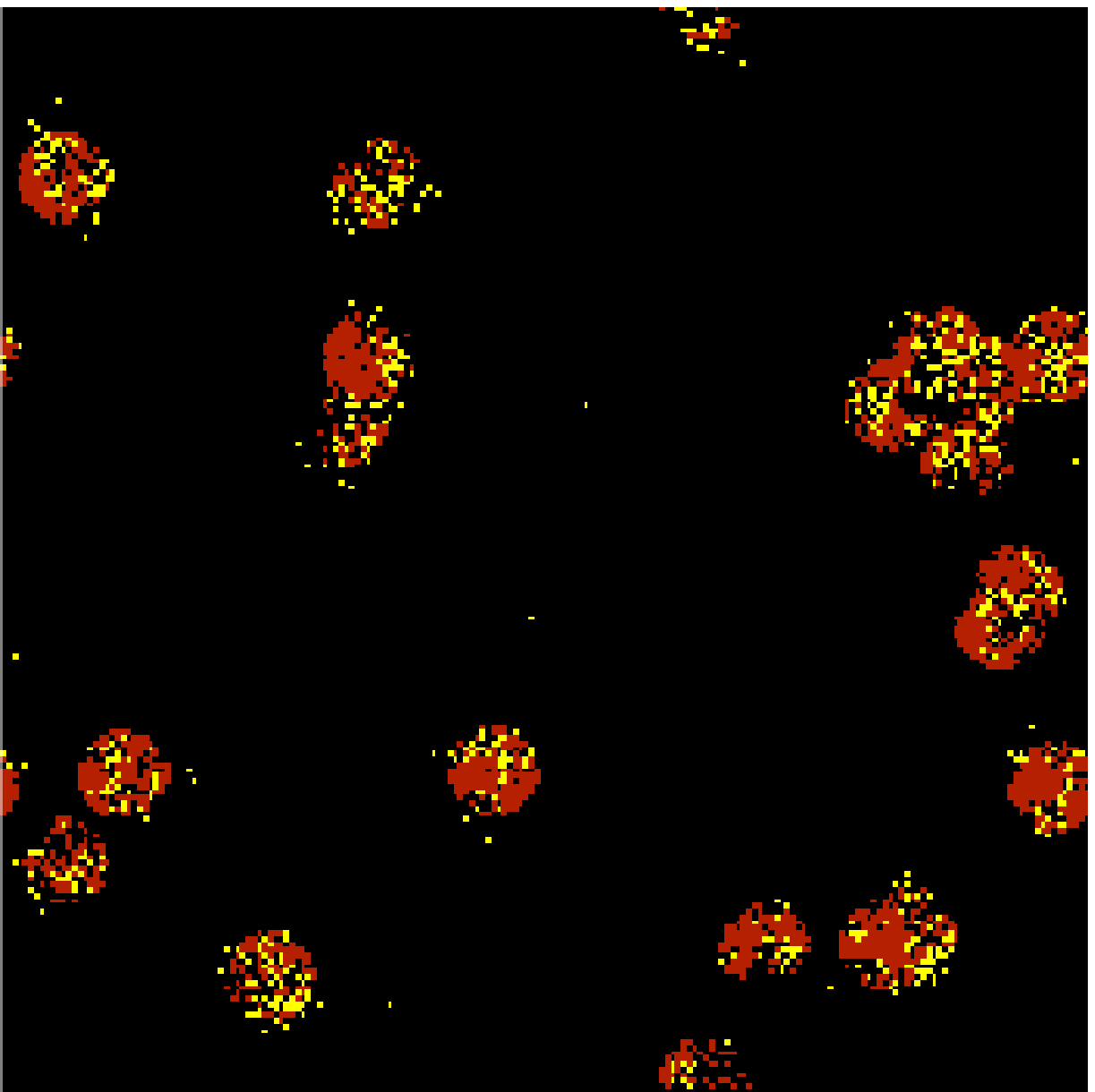}
}
\subfigure[ ]{
\includegraphics[scale=0.3]{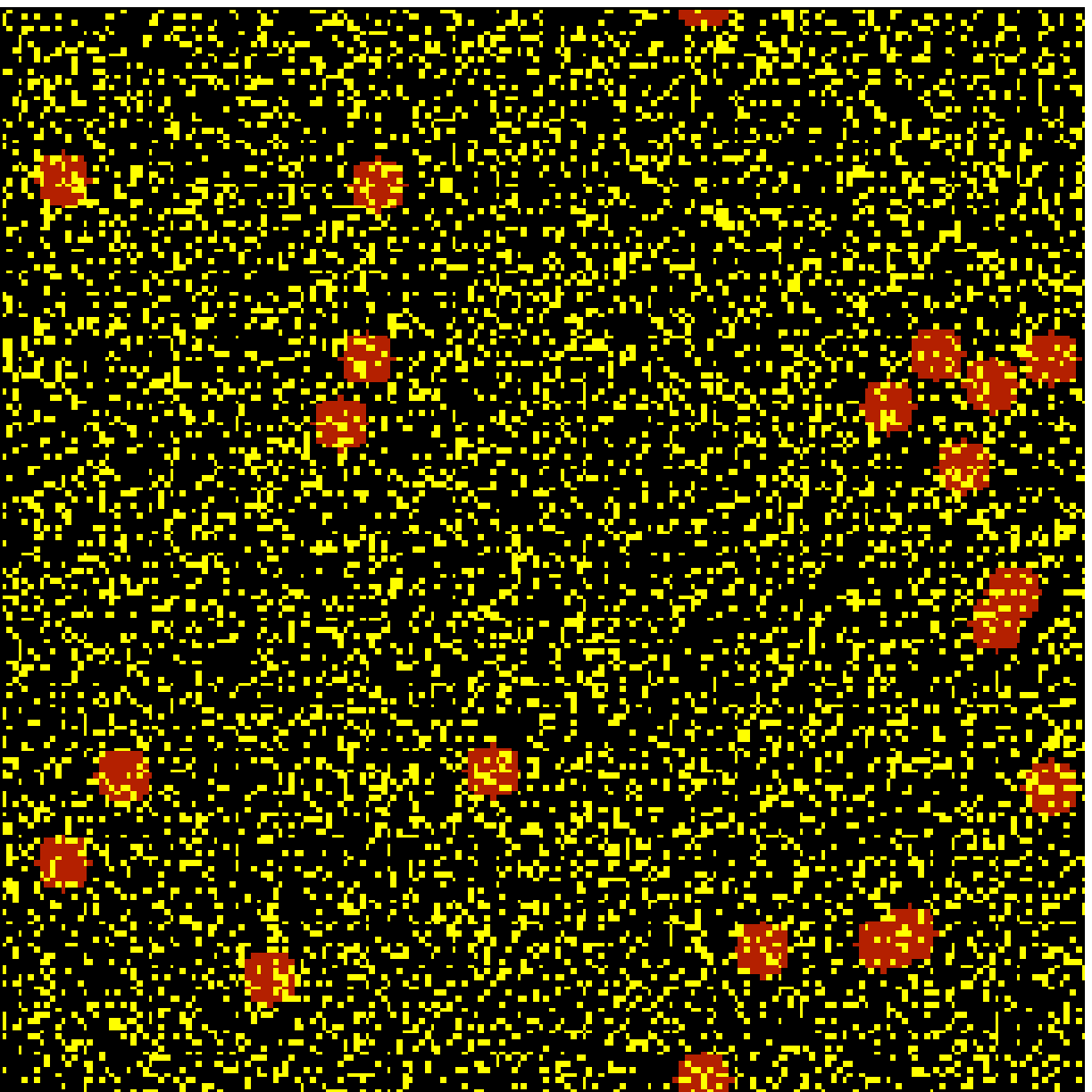} \includegraphics[scale=0.3]{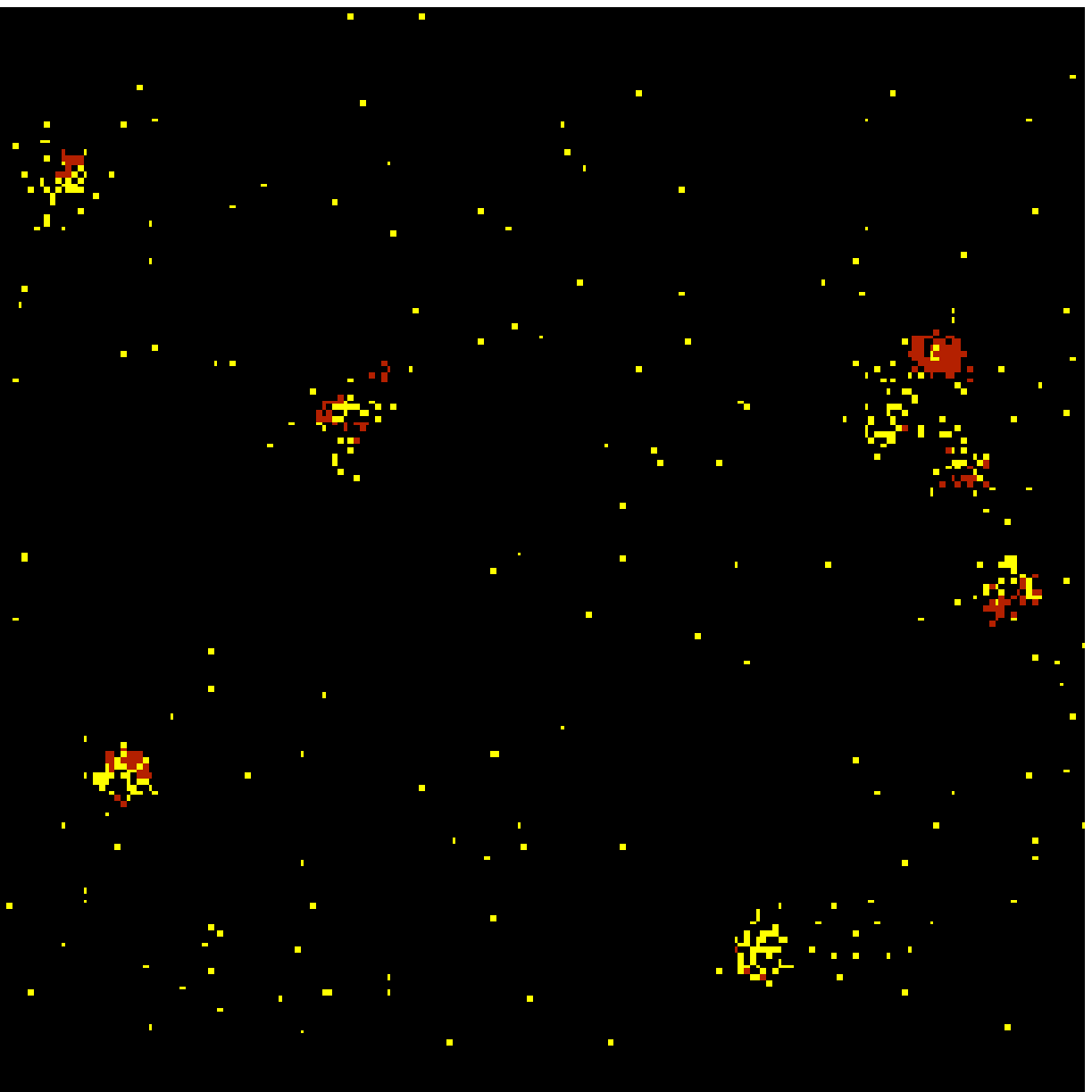}
}
\caption{Initial (left) and large-time (right) configurations of a metapopulation of Brownian predators (yellow dots) and randomly placed prey patches (prey are in red). (a) Same parameters as in Fig. \ref{brownbalcm}b (survival). (b) $\mu=0.08$ and smaller patch radii (global extinction by overexploitation, see movie SI2). The patches in the cases (a) and (b) have the same locations for easier comparison. $L=200$.} 
\label{brown}
\end{figure}

\subsection*{Results}
%
Highly-superdiffusive predators ($\beta\sim 1$) randomly sample space and therefore poorly exploit the patches. In the first scenario, their spatially average density is zero at large time, as expected from the mean-field analysis, whereas prey reach their maximum capacity in the patches (Figure \ref{brownbalcm}a). This situation corresponds to extinction by under-exploitation, see also movie SI1. With the same parameter values but Brownian mobility ($\beta>3$), in contrast, long-lived quasi-stationary states of predator-prey coexistence settle (Figure \ref{brownbalcm}b). Predator populations concentrate in the patches, as shown in a typical configuration (Figure \ref{brown}a): due to its slow diffusion, a Brownian predator located in a patch has a high probability to stay in its vicinity before dying, like its offsprings.

\begin{figure}
\centering
\includegraphics[scale=0.55]{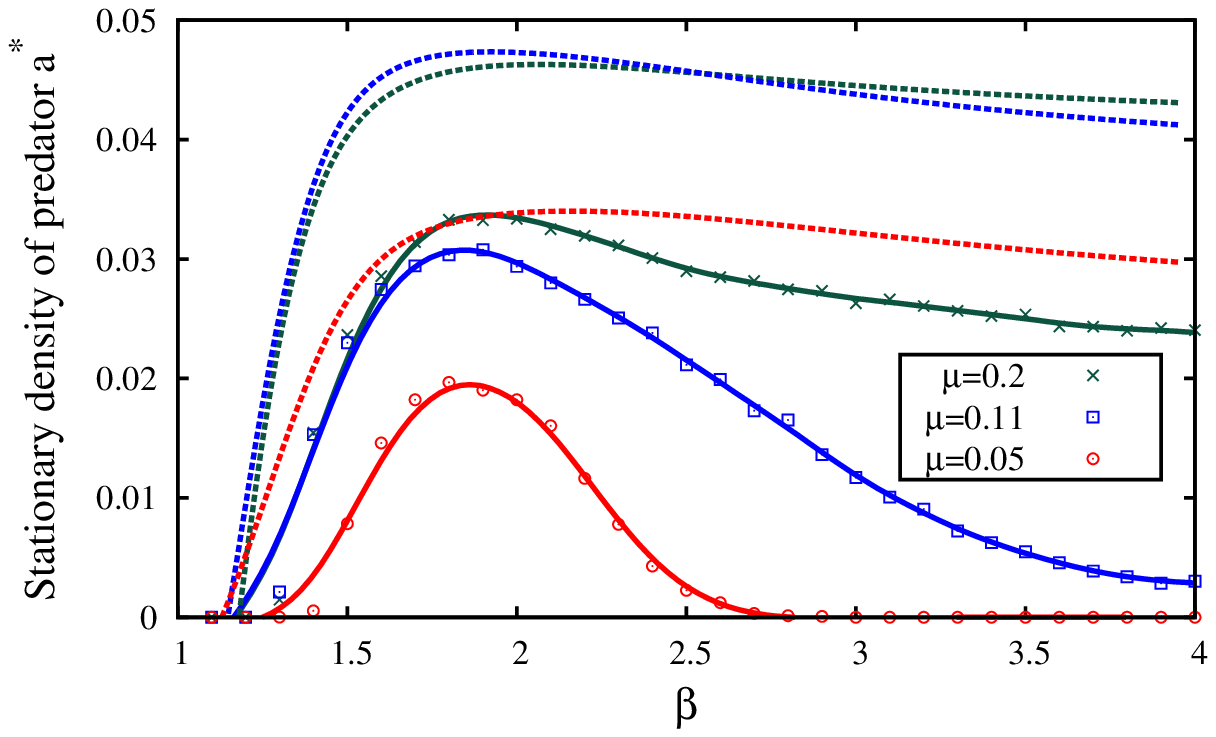}
\includegraphics[scale=0.55]{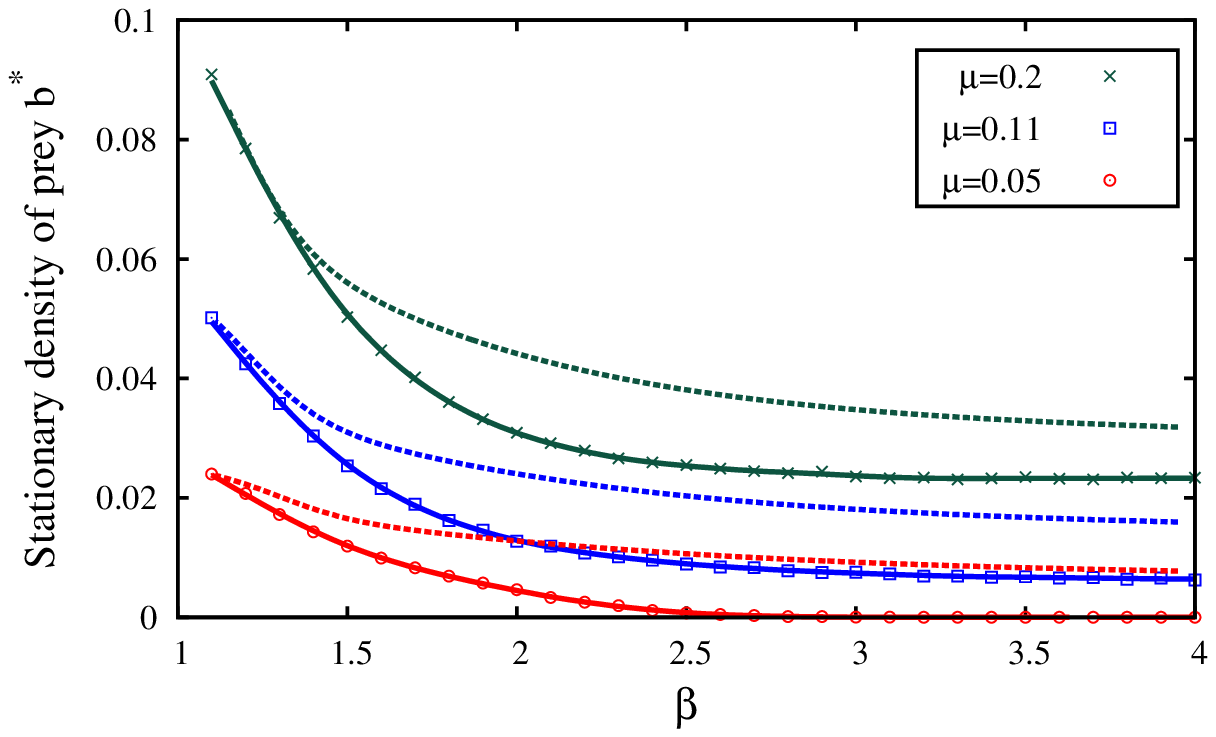}
\caption{(Top) Average predator density $a^*$ after $2000$ Monte Carlo steps as a function of $\beta$, and for 3 values of the mortality rate in the first scenario (solid lines). $L=500$ and the other parameters are those of Fig. \ref{brownbalcm}. At low $\mu$ (open red circles), $a^*$ is non-zero only in a relatively narrow region centered around $\beta \approx 1.8$. With dashed lines are shown the corresponding analytical calculations, Eqs. (\ref{Np})-(\ref{a0}). (Bottom) Average prey density $b^*$ obtained with the same parameters. See SI for more details.}
\label{avsbeta}
\end{figure}

The foregoing results suggest that Brownian motion in scarce and patchy environments stabilizes coexistence compared to the mean field expectation. However, such systems are not necessarily resilient in front of less favorable conditions. Figure \ref{brown}b illustrates a configuration where the predator mortality rate $\mu$ and the patch radius $R$, given by Eq. (\ref{R}), are lower than in Figure \ref{brown}a. Predators live longer and their number rapidly grows inside the patches, not letting the time for the prey to regenerate, see movie SI2. The patches are thus overexploited and irreversibly disappear after some unfavorable fluctuation (the empty patch is an absorbing state for the prey). Since the predators are left with no surrounding resources, they also go extinct. Figure \ref{avsbeta}-Top shows that the average density $a^*$ of normally diffusive predators (regime $\beta\ge3$) declines as $\mu$ decreases, and even vanishes when $\mu$ becomes too small. This important cause of extinction is not predicted by the analytic theory, which neglects temporal fluctuations.

Figure \ref{avsbeta}-Top shows that predators maximize their abundance when performing L\'evy flights with a particular exponent value, given by $\beta_c\approx 2$, all the other parameters being fixed (first scenario, see movie SI3). The location of the maximum depends little on the mortality rate $\mu$, as expected from the \lq\lq flat\rq\rq aspect of the theoretical curves of Fig. \ref{ranges}. In addition, less favorable conditions (lower mortality rate and smaller patches) mildly affect the average number of predators in the system when $\beta$ is around $\beta_c$ or below. In the Brownian case ($\beta\ge 3$), however, the same changes cause dramatic population declines as mentioned above. The predator population is not only maximal at $\beta_c$ but also persists if conditions are altered, a feature that ref. \cite{arnoldi} calls structural stability (the meaning of \lq\lq resilience” here). The movement strategy becomes crucial for the most fragile ecosystems (smallest values of $\mu$ in Fig. \ref{avsbeta}-top): predators face extinction due to under-exploitation or overexploitation depending on $\beta$, two situations which are avoided by adopting intermediate L\'evy flight strategies in a relatively narrow range around $\beta_c$. In such situations, L\' evy flights achieve a sustainable balance between exploitation and exploration, and are advantageous for stability and resilience.

As shown in Fig. \ref{avsbeta}-Top, abundances $a^*$ and $b^*$ given by the analytic theory (dotted lines) are in qualitative agreement with simulations. There are no adjustable parameters. Note however that theory significantly over-estimates $a^*$ and $b^*$, which do not vanish in the Brownian/low mortality regime. This is because local extinctions in the SLLVM are driven by fluctuations in finite size patches (where $b=0$ is an absorbing state) whereas noise is absent in the deterministic LV approach. Although less pronounced, the maximum of $a^*$ predicted by theory is in good agreement with simulations: from Eq. (\ref{opt}), we find $\beta_c\simeq 2.16$ ($\mu=0.05$), $1.92$ ($\mu=0.11$) and $2.06$ ($\mu=0.2$). 

Figure \ref{avsbeta}-Bottom displays the corresponding prey densities. Unlike $a^*$, $b^*$ decays monotonically with $\beta$ and is practically constant for $\beta\ge\beta_c$. Figs. \ref{avsbeta} illustrates the aforementioned resilience of predator populations with respect to changes in prey abundance: at $\beta=2$ the prey population decays by a factor of $2$ due to the change $\mu=0.2\rightarrow0.11$, whereas the predator population varies by less than $20\%$, therefore exhibiting a remarkable collective adaptation to the scarcer environment. Comparatively, for the same perturbation at $\beta=4$, where the number of prey are reduced by a factor of about $3$, the predator population is divided by $8$.

\begin{figure}
\centering
\subfigure[ ]{
\includegraphics[scale=.35]{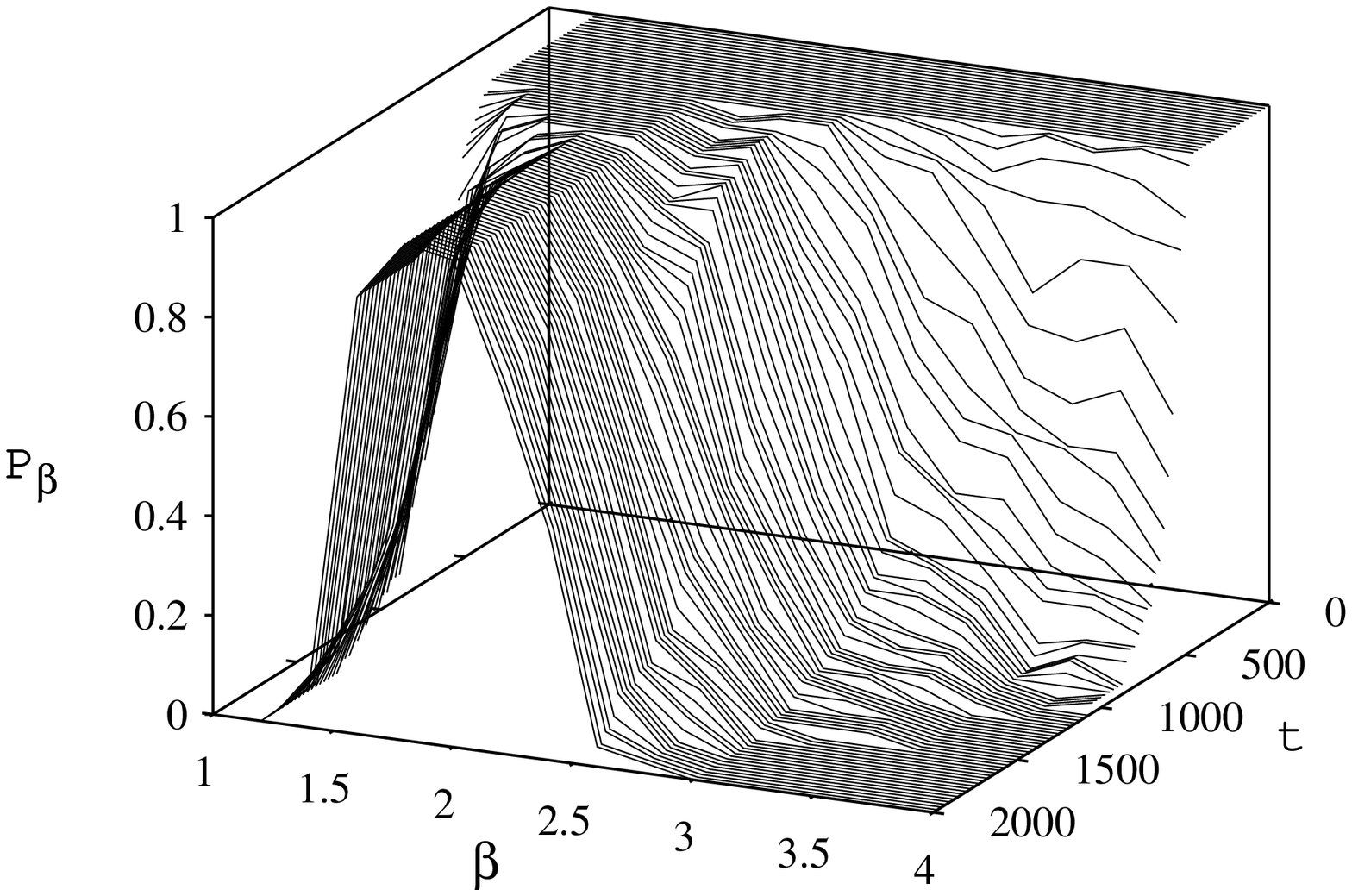}
}
\subfigure[ ]{ 
\includegraphics[scale=.5]{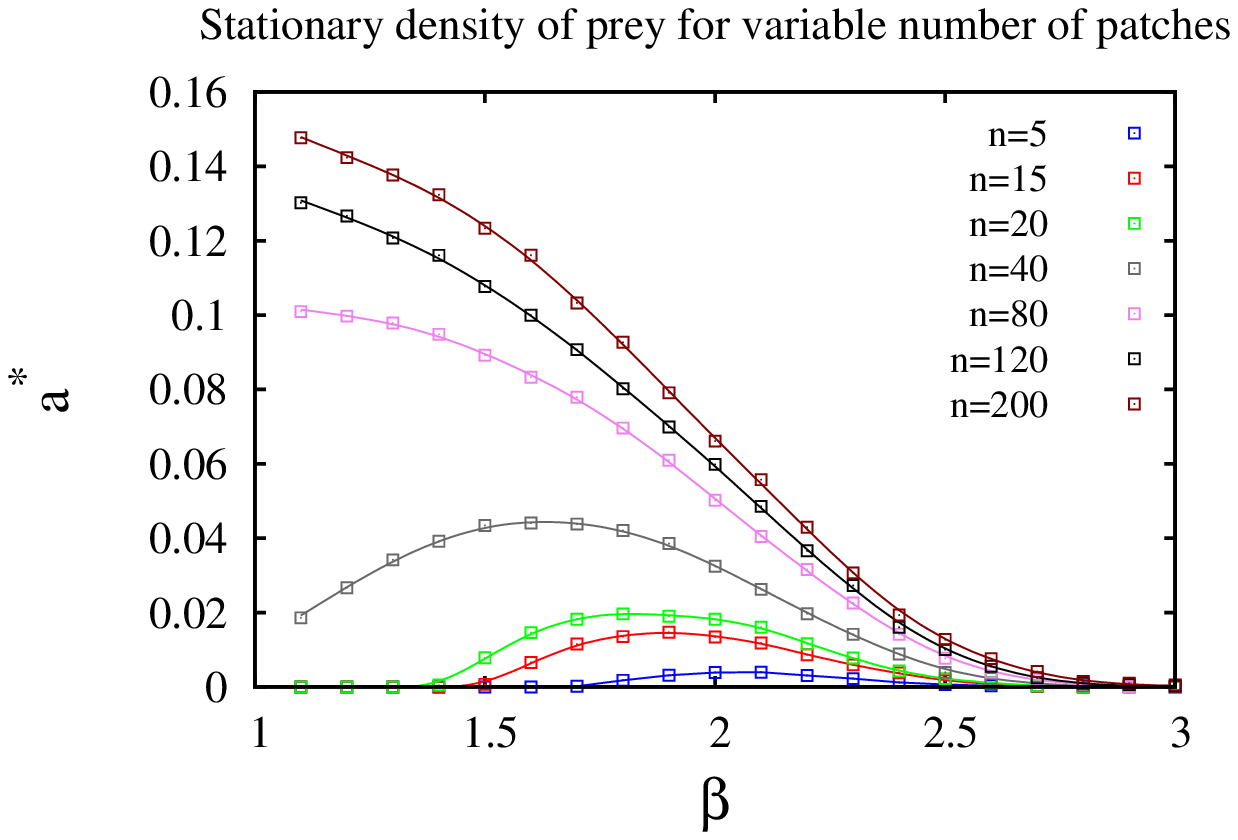}
}
\caption{(a) Evolution of the joint survival probability $P_{\beta}(t)$ up to $t=2000$, in unfavorable ecological conditions ($\sigma=0.2$, $\mu=0.05$, $n=20$, $L=200$, $R$ given by  Eq. \ref{R}). At large times, values of $\beta$ larger than 2.3 and lower than 1.5 result in a survival probability smaller than 0.5 due to overexploitation and underexploitation respectively. 
(b) Predator density as a function of the scaling exponent $\beta$ at fixed mortality rate ($\mu=0.05$) and patch radius ($R=4$), for different numbers of patches $n$ ($L=200$). Eq. (\ref{opt}) predicts without any adjustable parameters a maximum at $\beta_c\simeq 2.16$, $1.73$, $1.38$ for $n=20,$ $40$ and $80$, respectively, close to the simulation results ($\beta_c\simeq 1.85$, $1.65$, $1$).}
\label{evolucion}
\end{figure}

Another useful quantity in population dynamics is the joint survival probability, the probability that at least one individual of each species is alive at time $t$, denoted as $P_{\beta}(t)$. It is depicted in Figure \ref{evolucion}a. The  parameters in this example are chosen such that the system is subject to particularly unfavorable conditions for prey survival: small patches, low predator mortality rate and a lower prey recovering rate $\sigma$ than in Fig. \ref{avsbeta}. At large times, only a narrow range of value of $\beta$ around $2$ exhibits two-species coexistence ($P_{\beta}(t)\sim 1$).

We next vary the resource availability by means of the patch density, or $n$, keeping $\mu$ and $R$ fixed (scenario 2). When the mortality rate is low and prey is abundant ($n=200$ in Fig. \ref{evolucion}b), one could expect the predator abundance to be high, close to the mean field fixed point $(a^{(MF)},b^{(MF)})$, and to depend little on the movement strategy. However, Figure \ref{evolucion}b shows an example where no populations survive in the Brownian regime, while $a^*$ is non-vanishing in the L\'evy range. 

The foregoing finding is subtle and unexpected. The exponent $\beta_c$ that maximizes the predator population depends on the patch density. At small patch numbers, $\beta_c$  is in the vicinity of $2$ in this example and the predators do not survive if they perform other types of movements. As the patch density increases, the range of values of $\beta$ allowing predator survival increases and the optimal $\beta_c$ moves to the left until reaching unity. Importantly, the analytic expression (\ref{opt}) predicts this decrease of $\beta_c$ with patch density (see caption of Fig. \ref{evolucion}b). These results indicate that at low mortality, foraging strategies can be more flexible when resources are abundant, as long as predators avoid Brownian strategies. This fact could have profound evolutionary consequences.

\section*{Discussion}

In summary, simlpe population models with Lotka-Volterra interactions reveal that the stochastic movement strategies adopted by individuals searching for scarce resources have important consequences on the evolution of systems near extinction thresholds. These collective aspects cannot be directly inferred from single-forager random search models, which have been extensively studied \cite{gandhibook,intermsearch,Bar3,humphries2014optimal,bartulevin}. When resources are fragmented and regenerate slowly, predator metapopulations can avoid extinctions and maximize their abundance by means of L\'evy flights. 
For a wide range of demographic parameters, the multiple-scale structure of L\'evy mobility allows both local exploitation and long-range exploratory relocations that reduce the predation pressure on depleted zones. L\'evy populations are also resilient: a reduction of resources mildly affects their abundances, whereas it can produces rapid declines or extinctions when standard random-walk displacements are employed. In some cases, the range of random strategies allowing long-lived coexistence states becomes very narrow around the L\'evy exponent $\beta\approx 2$ as the patch density decreases.

Step-length distributions with exponents around 2 have been reported in many animal species \cite{bartu2003,RF,reynolds}, and also hunter-gatherers \cite{Brown} or fishing boats \cite{bertrand}. Our approach is useful for understanding aspects of human-environment interactions such as the multiple-scale displacements of fishing boats on the open ocean, where fish density is patchy and highly non-uniform \cite{makris}. These movements may result in a sustainable exploitation of fragile resources by giving profitable zones time to regenerate. Similar considerations apply to the nomadic hunter-gatherers of \cite{Brown}, who lived in resource-scarce lands. Future tests of our theory could also be performed in controlled laboratory experiments with micro-organisms like dinoflagellates, which are predators known to exhibit L\'evy patterns with exponent $\beta\simeq2$ at low prey concentrations \cite{bartu2003}. 

More generally, our results establish a connection between random search problems and the theory of metapopulations \cite{levins}, where a set of populations isolated in space becomes stabilized by fluxes between them. In our approach, L\'evy random motion effectively allows individuals born in a patch to visit other patches during their lifespan.
In a similar vein, power-law dispersal is known to increase asynchrony in  metapopulations with cyclic Rosenzweig-MacArthur or LV dynamics, making them less vulnerable \cite{gupta2017increased}. Further developments to many-species systems with realistic networks of trophic interactions \cite{mariana} and including heterogeneous patch size distributions are needed to study the effects of scale-free mobility on stability, sustainability and diversity. 

Traditionally, studies on animal foraging focus on the individual success for biological encounters such as the rate of prey capture. 
Our scope extends the notion of optimality in foraging by investigating the movement strategies that bring populations away from extinction thresholds. This is an essential step for developing movement-based ecological theories and concepts that could impact urgent problems in conservation biology.

\bibliographystyle{unsrt}
\bibliography{pnas-sample.bib}

\begin{thebibliography}{10}

\bibitem{valiente}
Alfonso Valiente-Banuet, Marcelo~A Aizen, Julio~M Alc{\'a}ntara, Juan Arroyo,
  Andrea Cocucci, Mauro Galetti, Mar{\'\i}a~B Garc{\'\i}a, Daniel Garc{\'\i}a,
  Jos{\'e}~M G{\'o}mez, Pedro Jordano, et~al.
\newblock Beyond species loss: the extinction of ecological interactions in a
  changing world.
\newblock {\em Functional Ecology}, 29(3):299--307, 2015.

\bibitem{pimm2006human}
Stuart Pimm, Peter Raven, Alan Peterson, {\c{C}}a{\u{g}}an~H
  {\c{S}}ekercio{\u{g}}lu, and Paul~R Ehrlich.
\newblock Human impacts on the rates of recent, present, and future bird
  extinctions.
\newblock {\em Proceedings of the National Academy of Sciences},
  103(29):10941--10946, 2006.

\bibitem{Barnosky}
Anthony~D Barnosky, Nicholas Matzke, Susumu Tomiya, Guinevere~OU Wogan, Brian
  Swartz, Tiago~B Quental, Charles Marshall, Jenny~L McGuire, Emily~L Lindsey,
  Kaitlin~C Maguire, et~al.
\newblock Has the earth/'s sixth mass extinction already arrived?
\newblock {\em Nature}, 471(7336):51--57, 2011.

\bibitem{DeVos}
Jurriaan~M De~Vos, Lucas~N Joppa, John~L Gittleman, Patrick~R Stephens, and
  Stuart~L Pimm.
\newblock Estimating the normal background rate of species extinction.
\newblock {\em Conservation Biology}, 29(2):452--462, 2015.

\bibitem{may}
Robert~McCredie May.
\newblock {\em Stability and complexity in model ecosystems}, volume~6.
\newblock Princeton university press, 2001.

\bibitem{solebascompte}
Jordi Bascompte and Ricard~V Sol{\'e}.
\newblock Habitat fragmentation and extinction thresholds in spatially explicit
  models.
\newblock {\em Journal of Animal Ecology}, pages 465--473, 1996.

\bibitem{pascual}
Ugo Bastolla, Miguel~A Fortuna, Alberto Pascual-Garc{\'\i}a, Antonio Ferrera,
  Bartolo Luque, and Jordi Bascompte.
\newblock The architecture of mutualistic networks minimizes competition and
  increases biodiversity.
\newblock {\em Nature}, 458(7241):1018--1020, 2009.

\bibitem{levins}
Richard Levins.
\newblock Some demographic and genetic consequences of environmental
  heterogeneity for biological control.
\newblock {\em American Entomologist}, 15(3):237--240, 1969.

\bibitem{Taylor}
LR~Taylor, IP~Woiwod, and JN~Perry.
\newblock The density-dependence of spatial behaviour and the rarity of
  randomness.
\newblock {\em The Journal of Animal Ecology}, pages 383--406, 1978.

\bibitem{Ritchie}
Mark~E Ritchie.
\newblock Scale-dependent foraging and patch choice in fractal environments.
\newblock {\em Evolutionary ecology}, 12(3):309--330, 1998.

\bibitem{Hamilton}
William~J Hamilton, William~M Gilbert, Frank~H Heppner, and Roy~J Planck.
\newblock Starling roost dispersal and a hypothetical mechanism regulating
  rhthmical animal movement to and from dispersal centers.
\newblock {\em Ecology}, 48(5):825--833, 1967.

\bibitem{Hamilton2}
William~J Hamilton and William~M Gilbert.
\newblock Starling dispersal from a winter roost.
\newblock {\em Ecology}, 50(5):886--898, 1969.

\bibitem{Okubo}
Akira Okubo and Smon~A Levin.
\newblock {\em Diffusion and ecological problems: modern perspectives},
  volume~14.
\newblock Springer Science \& Business Media, 2013.

\bibitem{hanski}
Ilkka Hanski and Oscar~E Gaggiotti.
\newblock {\em Ecology, genetics, and evolution of metapopulations}.
\newblock Academic Press, 2004.

\bibitem{niebuhr2015survival}
Bernardo~BS Niebuhr, Marina~E Wosniack, Marcos~C Santos, Ernesto~P Raposo,
  Gandhimohan~M Viswanathan, Marcos~GE Da~Luz, and Marcio~R Pie.
\newblock Survival in patchy landscapes: the interplay between dispersal,
  habitat loss and fragmentation.
\newblock {\em Scientific reports}, 5:11898, 2015.

\bibitem{may-miramontes}
MP~Hassell, O~Miramontes, P~Rohani, and RM~May.
\newblock Appropriate formulations for dispersal in spatially structured
  models: comments on bascompte \& sol{\'e}.
\newblock {\em Journal of Animal Ecology}, 64(5):662--664, 1995.

\bibitem{miramontes-rohani}
Pejman Rohani and Octavio Miramontes.
\newblock Host-parasitoid metapopulations: the consequences of parasitoid
  aggregation on spatial dynamics and searching efficiency.
\newblock {\em Proceedings of the Royal Society of London B: Biological
  Sciences}, 260(1359):335--342, 1995.

\bibitem{Tainaka}
K~Tainaka and Y~Itoh.
\newblock Topological phase transition in biological ecosystems.
\newblock {\em EPL (Europhysics Letters)}, 15(4):399, 1991.

\bibitem{Matsuda}
Hirotsugu Matsuda, Naofumi Ogita, Akira Sasaki, and Kazunori Sat{\=o}.
\newblock Statistical mechanics of population: the lattice lotka-volterra
  model.
\newblock {\em Progress of theoretical Physics}, 88(6):1035--1049, 1992.

\bibitem{Mobilia}
Mauro Mobilia, Ivan~T Georgiev, and Uwe~C Tauber.
\newblock Phase transitions and spatio-temporal fluctuations in stochastic
  lattice lotka-volterra models.
\newblock {\em J. Stat. Phys.}, 128, 2007.

\bibitem{maritan}
Rodrigo~P Rocha, Wagner Figueiredo, Samir Suweis, and Amos Maritan.
\newblock Species survival and scaling laws in hostile and disordered
  environments.
\newblock {\em Physical Review E}, 94(4):042404, 2016.

\bibitem{Lotka}
Alfred~J Lotka.
\newblock Analytical note on certain rhythmic relations in organic systems.
\newblock {\em Proceedings of the National Academy of Sciences}, 6(7):410--415,
  1920.

\bibitem{Volterra}
Vito Volterra.
\newblock Le{\c{c}}ons sur la th{\'e}orie math{\'e}matique de la lutte pour la
  vie.
\newblock {\em Gauthier-Villars, Paris}, 1936.

\bibitem{Bashan}
Amir Bashan, Travis~E Gibson, Jonathan Friedman, Vincent~J Carey, Scott~T
  Weiss, Elizabeth~L Hohmann, and Yang-Yu Liu.
\newblock Universality of human microbial dynamics.
\newblock {\em Nature}, 534(7606):259--262, 2016.

\bibitem{Perhar}
Gurbir Perhar, Noreen~E Kelly, Felicity~J Ni, Myrna~J Simpson, Andre~J Simpson,
  and George~B Arhonditsis.
\newblock Using daphnia physiology to drive food web dynamics: A theoretical
  revisit of lotka-volterra models.
\newblock {\em Ecological Informatics}, 35:29--42, 2016.

\bibitem{hanert}
Emmanuel Hanert.
\newblock Front dynamics in a two-species competition model driven by l{\'e}vy
  flights.
\newblock {\em Journal of theoretical biology}, 300:134--142, 2012.

\bibitem{morales}
Juan~Manuel Morales, Daniel~T Haydon, Jacqui Frair, Kent~E Holsinger, and
  John~M Fryxell.
\newblock Extracting more out of relocation data: building movement models as
  mixtures of random walks.
\newblock {\em Ecology}, 85(9):2436--2445, 2004.

\bibitem{nathan}
Ran Nathan, Wayne~M Getz, Eloy Revilla, Marcel Holyoak, Ronen Kadmon, David
  Saltz, and Peter~E Smouse.
\newblock A movement ecology paradigm for unifying organismal movement
  research.
\newblock {\em Proceedings of the National Academy of Sciences},
  105(49):19052--19059, 2008.

\bibitem{revilla2008individual}
Eloy Revilla and Thorsten Wiegand.
\newblock Individual movement behavior, matrix heterogeneity, and the dynamics
  of spatially structured populations.
\newblock {\em Proceedings of the National Academy of Sciences},
  105(49):19120--19125, 2008.

\bibitem{intermsearch}
Olivier B{\'e}nichou, C~Loverdo, M~Moreau, and R~Voituriez.
\newblock Intermittent search strategies.
\newblock {\em Reviews of Modern Physics}, 83(1):81, 2011.

\bibitem{benhamou}
Simon Benhamou.
\newblock Of scales and stationarity in animal movements.
\newblock {\em Ecology letters}, 17(3):261--272, 2014.

\bibitem{gandhibook}
Gandhimohan~M Viswanathan, Marcos~GE Da~Luz, Ernesto~P Raposo, and H~Eugene
  Stanley.
\newblock {\em The physics of foraging: an introduction to random searches and
  biological encounters}.
\newblock Cambridge University Press, 2011.

\bibitem{bartu2003}
Frederic Bartumeus, Francesc Peters, Salvador Pueyo, Celia Marras{\'e}, and
  Jordi Catalan.
\newblock Helical l{\'e}vy walks: adjusting searching statistics to resource
  availability in microzooplankton.
\newblock {\em Proceedings of the National Academy of Sciences},
  100(22):12771--12775, 2003.

\bibitem{RF}
Gabriel Ramos-Fern{\'a}ndez, Jos{\'e}L Mateos, Octavio Miramontes, Germinal
  Cocho, Hern{\'a}n Larralde, and Barbara Ayala-Orozco.
\newblock L{\'e}vy walk patterns in the foraging movements of spider monkeys
  (ateles geoffroyi).
\newblock {\em Behavioral ecology and Sociobiology}, 55(3):223--230, 2004.

\bibitem{boyermonos}
Denis Boyer, Gabriel Ramos-Fern{\'a}ndez, Octavio Miramontes, Jos{\'e}~L
  Mateos, Germinal Cocho, Hern{\'a}n Larralde, Humberto Ramos, and Fernando
  Rojas.
\newblock Scale-free foraging by primates emerges from their interaction with a
  complex environment.
\newblock {\em Proceedings of the Royal Society of London B: Biological
  Sciences}, 273(1595):1743--1750, 2006.

\bibitem{Atk}
RPD Atkinson, CJ~Rhodes, DW~Macdonald, and RM~Anderson.
\newblock Scale-free dynamics in the movement patterns of jackals.
\newblock {\em Oikos}, 98(1):134--140, 2002.

\bibitem{reynolds}
Andrew~M Reynolds, Alan~D Smith, Randolf Menzel, Uwe Greggers, Donald~R
  Reynolds, and Joseph~R Riley.
\newblock Displaced honey bees perform optimal scale-free search flights.
\newblock {\em Ecology}, 88(8):1955--1961, 2007.

\bibitem{Brown}
Clifford~T Brown, Larry~S Liebovitch, and Rachel Glendon.
\newblock L{\'e}vy flights in dobe ju/’hoansi foraging patterns.
\newblock {\em Human Ecology}, 35(1):129--138, 2007.

\bibitem{Sims}
David~W Sims, Emily~J Southall, Nicolas~E Humphries, Graeme~C Hays, Corey~JA
  Bradshaw, Jonathan~W Pitchford, Alex James, Mohammed~Z Ahmed, Andrew~S
  Brierley, Mark~A Hindell, et~al.
\newblock Scaling laws of marine predator search behaviour.
\newblock {\em Nature}, 451(7182):1098--1102, 2008.

\bibitem{deJager}
Monique de~Jager, Franz~J Weissing, Peter~MJ Herman, Bart~A Nolet, and Johan
  van~de Koppel.
\newblock L{\'e}vy walks evolve through interaction between movement and
  environmental complexity.
\newblock {\em Science}, 332(6037):1551--1553, 2011.

\bibitem{mirplosone}
Octavio Miramontes, Denis Boyer, and Frederic Bartumeus.
\newblock The effects of spatially heterogeneous prey distributions on
  detection patterns in foraging seabirds.
\newblock {\em PloS one}, 7(4):e34317, 2012.

\bibitem{humphries}
Nicolas~E Humphries, Henri Weimerskirch, Nuno Queiroz, Emily~J Southall, and
  David~W Sims.
\newblock Foraging success of biological l{\'e}vy flights recorded in situ.
\newblock {\em Proceedings of the National Academy of Sciences},
  109(19):7169--7174, 2012.

\bibitem{Bar3}
Frederic Bartumeus and J~Catalan.
\newblock Optimal search behavior and classic foraging theory.
\newblock {\em Journal of Physics A: Mathematical and Theoretical},
  42(43):434002, 2009.

\bibitem{bartulevin}
Frederic Bartumeus and Simon~A Levin.
\newblock Fractal reorientation clocks: Linking animal behavior to statistical
  patterns of search.
\newblock {\em Proceedings of the National Academy of Sciences},
  105(49):19072--19077, 2008.

\bibitem{prl2013}
Els Heinsalu, Emilio Hern{\'a}ndez-Garcia, and Crist{\'o}bal L{\'o}pez.
\newblock Clustering determines who survives for competing brownian and
  l{\'e}vy walkers.
\newblock {\em Physical Review Letters}, 110(25):258101, 2013.

\bibitem{Boyer2}
D~Boyer and O~L{\'o}pez-Corona.
\newblock Self-organization, scaling and collapse in a coupled automaton model
  of foragers and vegetation resources with seed dispersal.
\newblock {\em Journal of Physics A: Mathematical and Theoretical},
  42(43):434014, 2009.

\bibitem{bhat2017does}
U~Bhat, S~Redner, and O~B{\'e}nichou.
\newblock Does greed help a forager survive?
\newblock {\em Physical Review E}, 95(6):062119, 2017.

\bibitem{arnoldi}
Jean-Fran{\c{c}}ois Arnoldi and Bart Haegeman.
\newblock Unifying dynamical and structural stability of equilibria.
\newblock In {\em Proc. R. Soc. A}, volume 472, page 20150874. The Royal
  Society, 2016.

\bibitem{humphries2014optimal}
Nicolas~E Humphries and David~W Sims.
\newblock Optimal foraging strategies: L{\'e}vy walks balance searching and
  patch exploitation under a very broad range of conditions.
\newblock {\em Journal of theoretical biology}, 358:179--193, 2014.

\bibitem{bertrand}
Sophie Bertrand, Julian~M Burgos, Fran{\c{c}}ois Gerlotto, and Jaime Atiquipa.
\newblock L{\'e}vy trajectories of peruvian purse-seiners as an indicator of
  the spatial distribution of anchovy (engraulis ringens).
\newblock {\em ICES Journal of Marine Science}, 62(3):477--482, 2005.

\bibitem{makris}
Nicholas~C Makris, Purnima Ratilal, Deanelle~T Symonds, Srinivasan Jagannathan,
  Sunwoong Lee, and Redwood~W Nero.
\newblock Fish population and behavior revealed by instantaneous continental
  shelf-scale imaging.
\newblock {\em Science}, 311(5761):660--663, 2006.

\bibitem{gupta2017increased}
Anubhav Gupta, Tanmoy Banerjee, and Partha~Sharathi Dutta.
\newblock Increased persistence via asynchrony in oscillating ecological
  populations with long-range interaction.
\newblock {\em Physical Review E}, 96:042202, 2017.

\bibitem{mariana}
Cecilia Gonz{\'a}lez~Gonz{\'a}lez, Rafael L{\'o}pez~Mart{\'\i}nez, Sergio
  Hern{\'a}ndez~L{\'o}pez, and Mariana Ben{\'\i}tez.
\newblock A dynamical model to study the effect of landscape agricultural
  management on the conservation of native ecological networks.
\newblock {\em Agroecology and Sustainable Food Systems}, 40(9):922--940, 2016.

\end{thebibliography}

\end{document}